\documentclass[prl,reprint,amsmath,amssymb]{revtex4-1}
\usepackage{graphicx,color,amsmath,amssymb} 
\usepackage[colorlinks=true,urlcolor=blue, linkcolor=blue]{hyperref}

\def\({\left(}
\def\){\right)}

\def\d#1{#1^\dagger}

\newcommand\eq[1]{Eq.~(\ref{eq:#1})}
\newcommand\eqs[2]{Eqs.~(\ref{eq:#1},\ref{eq:#2})} 
\newcommand\fig[1]{Fig.~\ref{fig:#1}}
\newcommand\figs[2]{Figs.~\ref{fig:#1} \& \ref{fig:#2}}
\newcommand\bra[1]{\left\langle\,#1\,\right|} %\mid doesn't work with \left, \right
\newcommand\ket[1]{\left|\,#1\,\right\rangle}
\newcommand\scalprod[2]{\left\langle\,#1\,\right|\left.\,#2\,\right\rangle}
\newcommand\tr[2]{\mathrm{Tr}_{#1}\left[#2\right]}

\begin{document}

\title{On the experimental violation of Mermin's inequality with imperfect measurements}
\author{Ruffin Evans}
\author{Olivier Pfister}
\email{opfister@virginia.edu}
\affiliation{Department of Physics, University of Virginia, 382 McCormick Rd., Charlottesville, VA 22903, USA}

\begin{abstract}
 We investigate theoretically the feasibility of an experimental test of the Bell-type inequality derived by Mermin for correlated spins larger than ${1}/{2}$. Using the Schwinger representation, we link the output fields of two two-mode squeezers in order to create correlated effective spins between two observers. Spin measurements will be performed  by photon-number-resolved photodetection, which has recently come of age. We examine the effect of  nonideal detection quantum efficiency---and any other optical loss---on the violation margin of Mermin's inequality. We find that experimental violation is well accessible for spins larger than 1 for quantum efficiencies compatible with the current state of the art.
\end{abstract}

\maketitle

\onecolumngrid

\section{Introduction}        

The violation of Bell inequalities \cite{Bell1964} by measurements on quantum correlated systems with spins larger than the original ${1}/{2}$ value is of interest as a test of a new quantum regime, and possibly as a test of the quantum to classical transition \cite{Mermin1980}. In 1980, David Mermin showed that the ``standard'' Bell inequality cannot be violated for spins larger than ${1}/{2}$ and derived an alternative inequality which is violated by quantum mechanics for higher spins \cite{Mermin1980}. More recently, Collins, Gisin, Linden, Massar, and Popescu (CGLMP) discovered a generic method to investigate locality violations for higher-dimensional systems \cite{Collins2002}---Qdits rather than Qbits \cite{Mermin2003}---that may also offer increased violation and resistance to loss. 

Classic spin-${1}/{2}$ Bell inequalities have been violated under a variety of conditions over the past five decades. In particular, important experiments have closed various loopholes, such as the locality loophole \cite{Aspect1982,Aspect1982a,Weihs1998}, the fair sampling loophole \cite{Rowe2001}, and, most recently, the freedom-of-choice loophole \cite{Scheidl2010}.

In this paper, our motivation is to pave the way for an experimental test of Bell-type inequalities for spins larger than ${1}/{2}$ in general. An experimental violation of Bell inequalities for spin-1 systems was demonstrated \cite{Howell2002} but no other experimental work has been conducted on this subject to our knowledge. Here, we study  an experimental protocol that was proposed earlier by Reid and Drummond \cite{Drummond1983,Reid2002}, albeit in the context of dichotomic (Qbit) measurements. A related but different proposal was also made by Gerry and Albert \cite{Gerry2005}. This protocol (detailed in Section II) consists in using two pairs of entangled optical fields shared between two observers and viewed, in the Schwinger representation \cite{Schwinger1965}, as 2 spin $s$ particles in the zero-spin state used by Mermin's high-dimensional Bell inequality. Here, we are interested in the experimental feasibility of violating Mermin's inequality for $s>1$ and focus on a realistic model that includes arbitrary photodetection losses. This model is computationally intensive, hence we will focus solely on Mermin's inequality, which is the simplest computationally, even though an investigation of the CGLMP inequality (among others) is an equally interesting goal for future work. 

\subsection{An example}
Despite the relative simplicity of Bell's original inequality, it resists straightforward generalization to higher spins, as was pointed out by Mermin \cite{Mermin1980}. A good illustration of the situation is provided by examining the Clauser-Horne-Shimony-Holt (CHSH) inequality \cite{Clauser1969}, which can be constructed by taking the results of the measurements, by Alice and Bob respectively, of the components of two spins $\vec S_a$ and $\vec S_b$ along axes denoted by angles $\alpha$, $\beta$, $\gamma$, and $\delta$ about the $z$ axis \cite{Peres1995}:
\begin{align}
\mathcal C &= \left| m_{A,\alpha} + m_{A,\gamma}\right| m_{B,\beta} +  \left| m_{A,\alpha} - m_{A,\gamma}\right| m_{B,\delta}
\end{align}
For spins ${1}/{2}$, and normalizing the measurement results to values of $\pm 1$, it is easy to show that
\begin{equation}
\mathcal C \leqslant 2.
\end{equation}
However, as is well known, this counterfactual CHSH inequality is violated by quantum mechanics, which allows $\mathcal C_\mathit{max}=2\sqrt2$ for select angles when the spins are in a maximally entangled state. Let's now consider the general case of 2 spins $s$. It is relatively easy to show that, in this case, a counterfactual inequality is
\begin{align}
\mathcal C &\leqslant 2 s^2  
\end{align}
Like in the spin ${1}/{2}$ case, we now consider the quantum state of total spin 0
\begin{equation}
\ket{(ss)\,00} = \frac1{\sqrt{2s+1}} \sum_{m=-s}^s (-1)^{s-m} \ket{s, m}_a\ket{s, -m}_b,
\label{eq:0}
\end{equation}
for which the spin correlation matrix elements take the simple form \cite{Mermin1980}
\begin{equation}
\left\langle S_{A,\alpha}S_{B,\beta}\right\rangle  =  -\frac13 s(s+1) \cos(\alpha-\beta).
\label{eq:corr}
\end{equation}
In this case the CHSH inequality becomes
\begin{align}
 \left|\cos(\alpha-\beta)+\cos(\gamma-\beta)\right. %\nonumber\\
 +\left.\cos(\alpha-\delta)-\cos(\gamma-\delta)\right| \leqslant \frac{6s}{s+1}.
 \label{eq:schsh}  
\end{align}
%\begin{align}
% \left|\cos(\alpha-\beta)+\cos(\gamma-\beta)+\cos(\alpha-\delta)-\cos(\gamma-\delta)\right| \leqslant \frac{6s}{s+1}.  
%\end{align}
%\begin{widetext}
%\begin{align}
%\frac13 s(s+1) \left|\cos(\alpha-\beta)+\cos(\gamma-\beta)+\cos(\alpha-\delta)-\cos(\gamma-\delta)\right|\ \leqslant\ 2s^2.  
%\end{align}
%\end{widetext}
The absolute value on the left-hand side of \eq{schsh} is bounded from above by $2\sqrt2$, which corresponds to the maximum violation that quantum mechanics can offer. In that case, the inequality becomes
%\begin{equation}
%\frac{\sqrt2}3 \(1+\frac1s\) \leqslant 1,
%\end{equation}
\begin{equation}
s \geqslant \frac{\sqrt2}{3-\sqrt2} = 0.89,
\end{equation}
which is clearly satisfied by $s>1/2$. These spin values therefore pass the CHSH locality test.

However, this simple example should only be regarded as the tree that masks the forest. Indeed, there are many Bell-type inequalities that are violated by quantum mechanics for higher spins. Reference \cite{Howell2002} gives an instance for $s=1$, along with an experimental violation, and the aforementioned Refs.~\cite{Mermin1980,Collins2002} do so for arbitrary $s$. Moreover, work has also been done on Bell inequalities involving higher-dimensional systems  in terms of dichotomic-variable measurements (e.g.\ photon-counting thresholds, photon-number parity) \cite{Drummond1983,Gisin1992,Banaszek1998,Reid2002}.

\subsection{Mermin's inequality \cite{Mermin1980}}

We now recall Mermin's counterfactual inequality, which is violated by quantum mechanics in the zero eigenstate of the total spin:
\begin{equation}
s\langle|m_{A,\alpha}-m_{B,\beta}|\rangle_\mathit{av} \geqslant \langle m_{A,\alpha}m_{B,\gamma}\rangle_\mathit{av} + \langle m_{A,\beta}m_{B,\gamma}\rangle_\mathit{av}
\label{eq:Mermin}
\end{equation}
where
\begin{equation}
s\langle|m_{A,\alpha}-m_{B,\beta}|\rangle_\mathit{av} = \sum_{m,m'} |m-m'| P(m,m',\alpha-\beta).
\label{eq:diff}
\end{equation}
In the quantum mechanical derivation, the zero-spin state of \eq0 yields
\begin{equation}
P(m,m',\alpha-\beta) = \frac1{2s+1}\left|\bra me^{i(\alpha-\beta-\pi)S_y}\ket{m'}\right|^2
 = \frac{|d^s_{m m'}(-\alpha+\beta+\pi)|^2}{2s+1}
\end{equation}
and, as aforementioned in \eq{corr},
\begin{equation}
\langle m_{A,\alpha}m_{B,\gamma}\rangle_\mathit{av} = \left\langle S_{A,\alpha}S_{B,\beta}\right\rangle  =  -\frac13 s(s+1) \cos(\alpha-\beta),
\end{equation}
In evaluating the violation of \eq{Mermin}, Mermin computed (numerical) optimal measurement angles for maximal violation. The formulas above assume perfect measurements. In this work, we generalized Mermin's inequality to include the effect of measurement inefficiency in a particular optical implementation. The paper is organized as follows. In Section 2, we detail our experimental implementation proposal. In Section 3, we introduce our loss decoherence (detection quantum efficiency $\eta$) model and apply it to Mermin's inequality. In Section 4, we present the results of our numerical evaluations for $s=1/2$ to $9/2$ and $\eta=0.4$ to 1. We then conclude.

\section{Proposed experimental implementation}
\subsection{Einstein-Podolsky-Rosen (EPR) entanglement between Alice and Bob}

\begin{figure} [htbp]
%\vspace*{13pt}
%\centerline{\epsfig{file=fig1, width=8.2cm}} %100 percent
\centerline{\includegraphics[width=4.in]{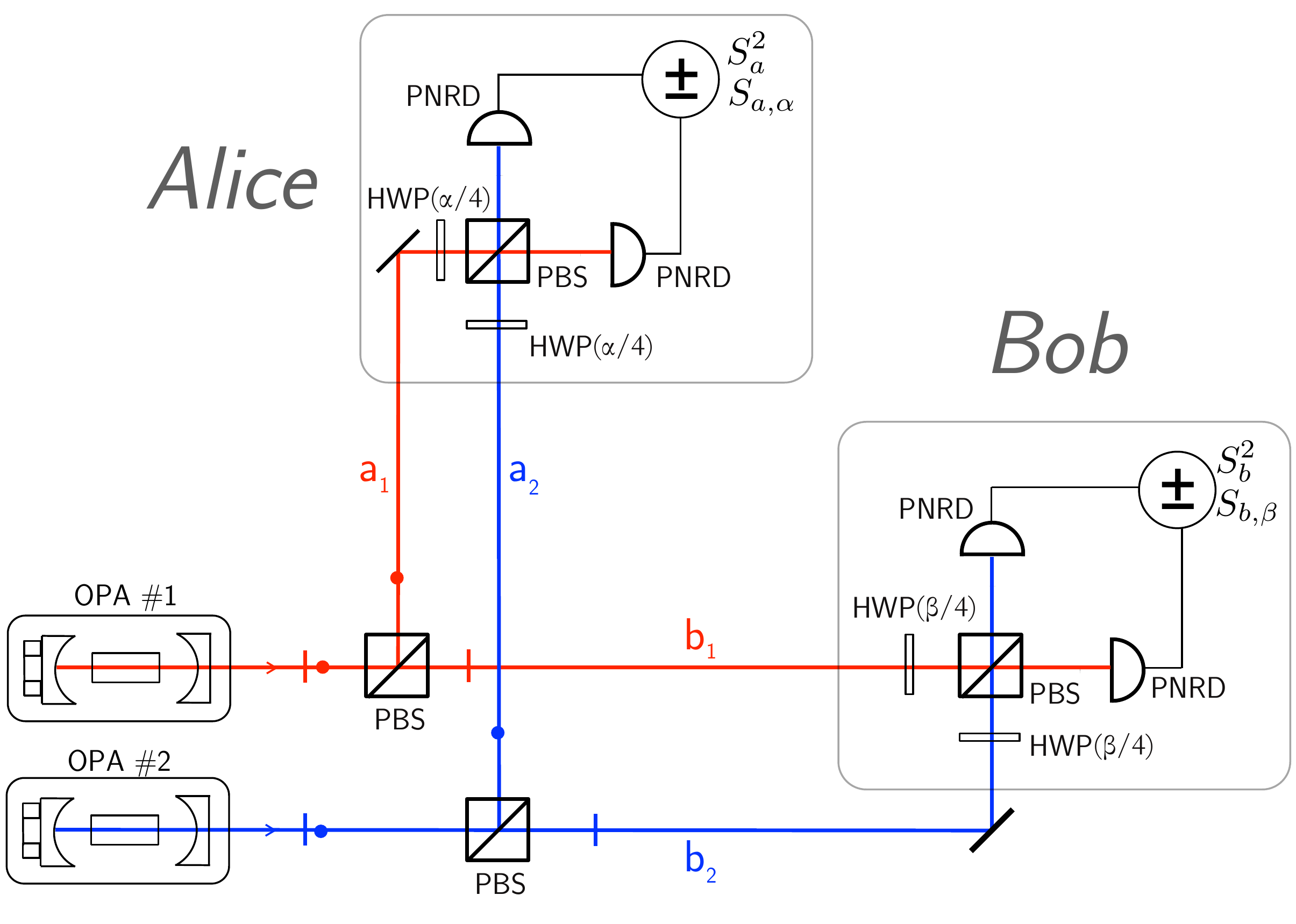}}
\vspace*{13pt}
\caption{\label{fig:setup}Proposed experimental setup for the violation of Mermin's inequality. OPA: optical parametric amplifier. PBS: polarizing beamsplitter. HWP: halfwave plate. PNRD: photon-number-resolving detector. Note that all optical paths in air between two consecutive PBS must be in phase, except for that of $a_2$ which must be exactly out of phase.}
\end{figure}

We consider the experimental setup sketched in \fig{setup}. Two two-mode squeezers, or optical parametric amplifiers OPA\#1 and OPA\#2,  are realized by using two optical parametric oscillators (OPOs) below threshold. 
%%%%%%%%%%%%%%%%%%%%%%%%%%%%%%%%%%%%%%%%%%%%%%%%%%%%%%%%%%%%%%%%%%
%%%%%%%%%%%%%%%%%%%%%%%%%%%%%%%%%%%%%%%%%%%%%%%%%%%%%%%%%%%%%%%%%%
Each OPA emits a two-mode squeezed, or EPR, state and each entangled mode is sent to each partner, Alice and Bob. Alice and Bob therefore each receive two independent fields, and each of their two fields is entangled with one of their partner's fields. This double EPR state is \cite{Walls1994}
\begin{align}
\ket{EPR^2} = &\sum_{n_1=0}^\infty  \frac{(\tanh r_1)^{n_1}}{\cosh r_1}\ket{n_1}_{A1}\ket{n_1}_{B1}%\nonumber\\
\otimes\sum_{n_2=0}^\infty  \frac{(\tanh r_2)^{n_2}}{\cosh r_2}\ket{n_2}_{A2}\ket{n_2}_{B2},
\end{align}
where $r_{1,2}$ are the squeezing parameters. We now show how this state can be recast as a perfectly entangled spin state, even for finite values of $r_{1,2}$. This was first pointed out by Drummond \cite{Drummond1983} and Reid et al. \cite{Reid2002}.

\subsection{Spin entanglement between Alice and Bob}
We use the Schwinger representation \cite{Schwinger1965}, which consists in defining the following effective spin operators
\begin{align}
S_{Ax} &= \frac12\(\d a_1a_2+a_1\d a_2\) \\
S_{Ay} &= \frac1{2i}\(\d a_1a_2-a_1\d a_2\) \\
S_{Az} &= \frac12\(\d a_1a_1-\d a_2a_2\) \\
S_A^2 &= \frac{\d a_1a_1+\d a_2a_2}2\(\frac{\d a_1a_1+\d a_2a_2}2+1\) 
\end{align}
for Alice, and
\begin{align}
S_{Bx} &= \frac12\(\d b_2b_1+b_2\d b_1\) \\
S_{By} &= \frac1{2i}\(\d b_2b_1-b_2\d b_1\) \\
S_{Bz} &= \frac12\(\d b_2b_2-\d b_1b_1\)\\
S_B^2 &= \frac{\d b_1b_1+\d b_2b_2}2\(\frac{\d b_1b_1+\d b_2b_2}2+1\)
\end{align}
for Bob. It is easy to check that these operators satisfy the canonical commutation relations of a quantum angular momentum. The usual eigenstates are those of $(S^2,S_{z})$: 
\begin{equation}
\ket{s_{A}, m_{A}}_A = \ket{\frac{n_{A1}+n_{A2}}2, \frac{n_{A1}-n_{A2}}2}_A \\
\label{eq:qnb}
\end{equation}
for Alice, and
\begin{equation}
\ket{s_{B}, m_{B}}_B = \ket{\frac{n_{B1}+n_{B2}}2, \frac{n_{B1}-n_{B2}}2}_B 
\end{equation}
for Bob. They can be measured in a straightforward manner if one can count photons with unity efficiency. The detectors in \fig{setup} will be photon-number-resolving superconducting transition-edge sensors, which have been installed in our laboratory at the University of Virginia, in  collaboration with the groups of Sae Woo Nam at the National Institute of Standards and Technology and Aaron Miller at Albion College. Such detectors have been shown to allow photon-number resolution at quantum efficiencies exceeding 95\% \cite{Lita2008}.

It is worthwhile to make explicit the physical meaning of these effective spin observables. Unsurprisingly, the total rotational energies given by observables $S_{A,B}^2$ coincide with the total energy in Alice's and Bob's fields, respectively. The spin projection along $z$ coincides with the photon number difference between the two fields on one partner's side. Finally, it is straightforward to see that the physical meaning of the spin projections along $x$ and $y$ is simply the phase difference between the two fields on one partner's side, since the observables $S_{x,y}$ are clearly interference terms. These considerations therefore completely specify the measurement procedure \cite{Kim1998}, which we detail in the next subsection.

We can thus rewrite the double EPR state as
%\begin{widetext}
\begin{equation}
\ket{EPR^2} = \sum_{s=0}^\infty\sum_{m=-s}^s  \frac{(\tanh r_1)^{s+m}}{\cosh r_1}\frac{(\tanh r_2)^{s-m}}{\cosh r_2}\ket{s\ m}_A\ket{s\ -m}_B.
\label{eq:EPR2}
\end{equation}
%\end{widetext}
This state can be put in an interesting form provided the squeezing parameters are identical $r_1=r_2=r$. We then obtain 
\begin{equation}
\ket{EPR^2} = \sum_{s=0}^\infty \left[\frac{(\tanh r)^s}{\cosh r}\right]^2
\sum_{m=-s}^s  \ket{s\ m}_A\ket{s\ -m}_B.
\end{equation}
This state is clearly a {maximally} entangled spin state at any given value of $s$, which can be heralded or defined by the measurement of the total photon number on Alice's and Bob's sides. In addition, we can obtain the exact zero-total-spin eigenstate by phase shifting either the $A2$ or the $B2$ optical path by $\pi$: 
\begin{equation*}
e^{-i\pi N_{A2}}\ket{n_2}_{A_2} = e^{-i\pi n_2}\ket{n_2}_{A_2} = (-1)^{n_2}\ket{n_2}_{A_2},
\end{equation*}
which yields
\begin{equation}
\ket{EPR^2} = \sum_{s=0}^\infty \sqrt{2s+1}\left[\frac{(\tanh r)^s}{\cosh r}\right]^2
\ket{(ss)00}.
\end{equation}
Thus, the double EPR state can be made equivalent to the eigenstate of two spins $s$ coupled into a zero total spin. Do note that the maximal entanglement of the state is solely guaranteed by $r_1=r_2=r$, and in no way by the value of $r$, which does not have, in particular, to be large.

\subsection{Spin measurement procedure}

In order to rotate the ``spin analyzers'' on Alice's and Bob's sides, we need to be able to measure the arbitrary spin component defined as 
\begin{equation}
S_\alpha = \cos\alpha\ S_z +\sin\alpha\ S_x.
\label{eq:alpha}
\end{equation}
This is the purpose of the halfwave plates and subsequent polarizing beam splitters (PBS) in \fig{setup}, which implement variable beamsplitters right before each set of detectors and can therefore rotate the measurement of the effective spin component from an intensity difference ($S_z$) measurement to an interference (say, $S_x$) one. Indeed, if both waveplates are at the same angle $\alpha/4$ with respect to the PBS, the fields at each of Alice's detectors will be
\begin{align}
A_1 & = \cos\frac\alpha2\ a_1+\sin\frac\alpha2\ a_2  \\
A_2 & = -\sin\frac\alpha2\ a_1+\cos\frac\alpha2\ a_2  
\end{align}
and the expressions of the photon number on each of Alice's detectors will be
\begin{align}
N_{A1} &= \cos^2\frac\alpha2\ \d a_1a_1 +\sin^2\frac\alpha2\ \d a_2a_2 + \frac{\sin\alpha}2\(\d a_1a_2+a_1\d a_2\) \\
N_{A2} &= \sin ^2\frac\alpha2\ \d a_1a_1 + \cos^2\frac\alpha2\ \d a_2a_2 - \frac{\sin\alpha}2\(\d a_1a_2+a_1\d a_2\) 
\end{align}
so that the sum and difference measurements give
\begin{align}
N_{A+} &= \d a_1a_1 +\d a_2a_2 \qquad \quad \leftrightarrow\ \ \qquad S_A^2\\
N_{A-} &= 2\cos\alpha\ S_{Az} + 2\sin\alpha\ S_{Ax}  \equiv 2 S_{A,\alpha}
\end{align}
and the same of course on Bob's side. Hence, this experimental setup realizes the  spin measurements of \eq{alpha} for testing Mermin's inequality. 

We now turn to the core of this paper, i.e.\  the investigation of nonideal spin measurements, as defined by the nonideal photon-counting which stems from optical losses in the experiment.

\section{Testing Mermin's inequality in the presence of losses}

\subsection{Modeling loss decoherence}

We model loss decoherence by considering the effect of beam splitters of transmissivity $\eta_j$, $j\in[1,4]$, on the detected crosscorrelations. These beamsplitters can be assumed to be just before each detector but their  position  in the experiment between the source and the detector doesn't actually matter. They just serve as---valid---models \cite{Yuen1978,Yuen1980} for compounded propagation and detection losses. 

The situation is complicated by the fact that decoherence turns the pure state $\ket\phi$ into a statistical mixture. We must therefore use a density operator description from now on. The statistical mixture is obtained from the density operator,  partially traced over the loss ports of the beamsplitter:
\begin{equation}
\rho' = \tr{\mathit{loss\, ports}}{U_{BS}\ket\phi\bra\phi\d U_{BS}},
\label{eq:decoh}
\end{equation}
where $U_{BS}$ is the unitary operator describing the action of 4 beam splitters, each with two input ports (one of which occupied by a vacuum field) and two output ports [one of which is traced over in \eq{decoh}]. 
%The probability analog to the left-hand side of \eq{Mermin} is then
%\begin{equation}
%P(m,m',\widehat{\hat n,\hat n'}) = \tr{}{\rho'\ket m_{\hat n}\ket{m'}_{\hat n'}\bra m_{\hat n}\bra{m'}_{\hat n'}}.
%\end{equation}
%and the crosscorrelations on the right-hand side are 
We now turn to the detailed expression of $\rho'$. 

\subsubsection{Decohered Fock state} 

Modeling the effect of losses as beamsplitters of transmissivity $\eta$, we can use the well-known result \cite{Caves1980} that a Fock state interfering with vacuum at a beamsplitter is turned into the ``binomially entangled'' output
\begin{equation}
U_{BS}\ket n_a\ket0_{a'} = \sum_{k=0}^n %\binom
\binom nk^\frac12 {\eta^\frac k2(1-\eta)^\frac{n-k}2} \ket k_b\ket{n-k}_{b'}
\end{equation}
In the case of decoherence, only one output port is available and the other must be traced over. The initial pure-state mode $a$ is therefore turned into a statistical mixture described by 
\begin{equation}
\rho_\eta 
=  \tr{b'}{U_{BS}(\eta)\ket n_a\ket0_{a'}\bra 0_{a'}\bra n_a\d U_{BS}(\eta)}.
\end{equation}

Evaluating this expression yields the familiar binomial probability distribution for the output photon number
\begin{align}
%\rho &= \sum_l \sum_{k=0}^n\sum_{k'=0}^n \binom nk^\frac12\binom n{k'}^\frac12 \eta^\frac{k+k'}2(1-\eta)^{n-\frac{k+k'}2} \ket k_b\bra {k'}\scalprod l{n-k}_{b'}\scalprod{n-k'}l_{b'}\\
\rho_\eta&= \sum_{k=0}^n%\binom 
\binom nk \eta^k(1-\eta)^{n-k} \ket k_b\bra {k}.
\end{align}

\subsubsection{Decohered arbitrary single-mode  state} 

We now consider any superposition of Fock states $\ket\psi=\sum_n \psi_n\ket n$. The initial density operator is thus 
\begin{equation}
\ket\psi\bra\psi=\sum_{n,n'} \psi_n\psi^*_{n'}\ket n\bra{n'}
\end{equation}
and we can show, following the exact same procedure as before, that the partial trace associated with the decoherence process turns the operator $\ket n\bra{n'}$ into the operator
\begin{align}
%&\ket n\bra{n'} \mapsto \nonumber\\
\Pi_{nn'} = &\sum_l \sum_{k=0}^n\sum_{k'=0}^n \binom nk^\frac12\binom {n'}{k'}^\frac12 \eta^\frac{k+k'}2(1-\eta)^\frac{n-k+n'-k'}2 \ket k_b\bra {k'}\scalprod l{n-k}_{b'}\scalprod{n'-k'}l_{b'}\\
=& \sum_{k=\max(0,n-n')}^n\binom nk^\frac12\binom {n'}{k+n'-n}^\frac12 \eta^{k+\frac{n'-n}2}(1-\eta)^{n-k} \ket k_b\bra {k+n'-n},
\end{align}
an expression which we can make more symmetric by using the sum index change $ \kappa=k+(n'-n)/2$ and the fact that $\max[\pm (n-n')] \equiv |n-n'|$:
\begin{equation}
\Pi_{nn'}  = \sum_{\kappa=\frac{|n-n'|}2}^{\frac{n+n'}2}\binom n{\kappa+\frac{n-n'}2}^\frac12\binom {n'}{\kappa+\frac{n'-n}2}^\frac12 \eta^{\kappa}(1-\eta)^{-\kappa+\frac{n+n'}2} \ket {\kappa+\frac{n-n'}2}_b\bra {\kappa+\frac{n'-n}2}.
\label{eq:decohsingle}
\end{equation}

\subsubsection{Decohered arbitrary two-mode  state} 
We now consider an arbitrary two-mode state
\begin{align}
\ket\psi\bra\psi
&=\sum_{n_1,n_2,n'_1,n'_2} \psi_{n_1,n_2}\psi ^*_{n'_1,n'_2}\ket{n_1}\bra{n'_1}\otimes\ket{n_2}\bra{n'_2}
\end{align}
The action of decoherence is independent on each mode, therefore we can directly use the result of \eq{decohsingle} to show that the operator $\ket{n_1}\bra{n'_1}\otimes\ket{n_2}\bra{n'_2}$ becomes
\begin{align}
&\Pi_{n_1n_1'n_2n_2'} =\nonumber\\
&\sum_{\kappa_1=\frac{|n_1-n'_1|}2}^{\frac{n_1+n'_1}2}
\binom {n_1}{\kappa_1+\frac{n_1-n'_1}2}^\frac12
\binom {n'_1}{\kappa_1+\frac{n'_1-n_1}2}^\frac12 
\eta_1^{\kappa_1}
(1-\eta_1)^{-\kappa_1+\frac{n_1+n'_1}2} 
\textstyle
\ket {\kappa_1+\frac{n_1-n'_1}2}_1\bra {\kappa_1+\frac{n'_1-n_1}2}\nonumber\\
&\otimes\sum_{\kappa_2=\frac{|n_2-n'_2|}2}^{\frac{n_2+n'_2}2}
\binom {n_2}{\kappa_2+\frac{n_2-n'_2}2}^\frac12
\binom {n'_2}{\kappa_2+\frac{n'_2-n_2}2}^\frac12 
\eta_2^{\kappa_2}
(1-\eta_2)^{-\kappa_2+\frac{n_2+n'_2}2} 
\textstyle
\ket {\kappa_2+\frac{n_2-n'_2}2}_2\bra {\kappa_2+\frac{n'_2-n_2}2}. 
\label{eq:2mode}
\end{align}
Now that we have two field modes, we can make use of the Schwinger representation to understand the effects of optical losses on the effective spins that will be used in the final experiment. We find:
\begin{align}
\ket\psi\bra\psi
=\sum_{s,m,s',m'} \psi_{s,m}\, \psi ^*_{s',m'}\ket{s\ m}\bra{s'\ m'},
\label{eq:schwing}
\end{align}
with the same definitions as \eq{qnb} for the spin numbers. In addition, we replace the summation indices $\kappa_j$ in \eq{2mode} with the spin variable $(\sigma,\mu)$ such that
\begin{align}
\sigma &= \frac12\(\kappa_1+\frac{n_1-n'_1}2+\kappa_2+\frac{n_2-n'_2}2\) \\
\mu &= \frac12\(\kappa_1+\frac{n_1-n'_1}2-\kappa_2-\frac{n_2-n'_2}2\) 
\end{align}
These expressions define both kets in \eq{2mode} as the spin ket $\ket{\sigma,\mu}$ and it is straightforward to show that the bras become the spin bra $\bra{\sigma+s'-s,\mu+m'-m}$. Other helpful relations are
\begin{align}
\kappa_1 &= \sigma+\mu +\frac{s'-s+m'-m}2 \\
\kappa_2 &= \sigma-\mu +\frac{s'-s-m'+m}2. 
\end{align}
The upper bound for $\sigma$ is $s$ and the lower bound depends on the relative numbers of photons $n_1$ and $n_2$, as given in Table 1. 
%\vspace*{4pt}   %only when needed

\begin{table}[hb]
\caption{Lower bounds $\sigma_\mathit{min}$ in \eq{DecohdSpin}.}
\centerline{
\begin{tabular}{c | c }
Case & $\sigma_\mathit{min}$ \\
\hline
&\\
$n_1\geqslant n_2$ and $n'_1\geqslant n'_2$ & $s-s'$ \\
&\\
$n_1\leqslant n_2$ and $n'_1 \leqslant n'_2$ & 0 \\
&\\
$n_1\geqslant n_2$ and $n'_1 \leqslant n'_2$ & $\frac{s-s'+m-m'}2$ \\
&\\
$n_1 \leqslant n_2$ and $n'_1\geqslant n'_2$ & $\frac{s-s'-m+m'}2$ \\
&\\
\hline
\end{tabular}}
\end{table}

%If $n_1\geqslant n_2$ and $n'_1\geqslant n'_2$, $\sigma_\mathit{min}=s-s'$. 
%If $n_1\leqslant n_2$ and $n'_1 \leqslant n'_2$, $\sigma_\mathit{min}=0$. 
%If $n_1\geqslant n_2$ and $n'_1 \leqslant n'_2$, $\sigma_\mathit{min}=(s-s'+m-m')/2$. 
%If $n_1 \leqslant n_2$ and $n'_1\geqslant n'_2$, $\sigma_\mathit{min}=(s-s'-m+m')/2$.  
Then, the action of decoherence turns operator  $\ket{s\ m}\bra{s'\ m'}$ in \eq{schwing} into the operator
\begin{align}
\Pi_{s ms' m'} = 
&
\sum_{\sigma=\sigma_\mathit{min}}^s\sum_{\mu=-\sigma}^\sigma
\eta_1^{\sigma+\mu+\frac{s'-s+m'-m}2}\eta_2^{\sigma-\mu+\frac{s'-s-m'+m}2}
(1-\eta_1)^{-\sigma-\mu+s+m} 
(1-\eta_2)^{-\sigma+\mu+s-m} \nonumber\\
&
\binom {s+m}{\sigma+\mu}^\frac12
\binom {s'+m'}{\sigma+\mu+s'-s+m'-m}^\frac12
\binom {s-m}{\sigma-\mu}^\frac12
\binom {s'-m'}{\sigma-\mu+s'-s-m'+m}^\frac12 \nonumber\\
&
\ket {\sigma,\mu}\bra {\sigma+s'-s,\mu+m'-m} 
\label{eq:DecohdSpin}
\end{align}
The expression in \eq{DecohdSpin} is thus Alice's decohered $\ket{s\ m}\bra{s'\ m'}$ operator. 
Note that all $s,m,s',m'$ must be the same for Alice and Bob because of the quantum EPR correlations before decoherence (and, again, the finiteness of squeezing plays no role in this). This yields the final decohered entangled density operator, which will be used to compute all averages in Mermin's inequality:
\begin{align}
\rho' = &\sum_{s=0}^\infty\sum_{s'=0}^\infty\tau_{sr}^2\tau_{s'r}^2\sum_{m=-s}^s\sum_{m'=-s'}^{s'} (-1)^{s-m+s'-m'}
\sum_{\sigma_a=\sigma_\mathit{min}}^s\sum_{\mu_a=-\sigma_a}^{\sigma_a}
\sum_{\sigma_b=\sigma_\mathit{min}}^s\sum_{\mu_b=-\sigma_b}^{\sigma_b}\nonumber\\
& \quad
\eta_{a1}^{\sigma_a+\mu_a+\frac{s'-s+m'-m}2}\eta_{a2}^{\sigma_a-\mu_a+\frac{s'-s-m'+m}2}
(1-\eta_{a1})^{-\sigma_a-\mu_a+s+m} 
(1-\eta_{a2})^{-\sigma_a+\mu_a+s-m} \nonumber\\
& \quad
\binom {s+m}{\sigma_a+\mu_a}^\frac12
\binom {s'+m'}{\sigma_a+\mu_a+s'-s+m'-m}^\frac12
\binom {s-m}{\sigma_a-\mu_a}^\frac12
\binom {s'-m'}{\sigma_a-\mu_a+s'-s-m'+m}^\frac12 \nonumber\\
& \quad
\ket {\sigma_a,\mu_a}_A\bra {\sigma_a+s'-s,\mu_a+m'-m} \nonumber\\
& \quad
\eta_{b1}^{\sigma_b+\mu_b+\frac{s'-s+m'-m}2}\eta_{b2}^{\sigma_b-\mu_b+\frac{s'-s-m'+m}2}
(1-\eta_{b1})^{-\sigma_b-\mu_b+s+m} 
(1-\eta_{b2})^{-\sigma_b+\mu_b+s-m} \nonumber\\
& \quad
\binom {s+m}{\sigma_b+\mu_b}^\frac12
\binom {s'+m'}{\sigma_b+\mu_b+s'-s+m'-m}^\frac12
\binom {s-m}{\sigma_b-\mu_b}^\frac12
\binom {s'-m'}{\sigma_b-\mu_b+s'-s-m'+m}^\frac12 \nonumber\\
& \quad
\ket {\sigma_b,\mu_b}_B\bra {\sigma_b+s'-s,\mu_b+m'-m}, 
\end{align}
where $\tau_{sr}=(\tanh r)^n/\cosh r$.

\subsection{Computing Mermin's inequality in the presence of losses}

In order to evaluate the inequality in \eq{Mermin}, we derive the expressions of the dual-measurement probability (left-hand side term) and of the crosscorrelations (right-hand side term).

\subsubsection{Dual-measurement probability}

We now derive the probability distribution of measurement results $(s_a, m_a)$ and $(s_b, m_b)$, with respective ``analyzer'' directions $\alpha$ and $\beta$, in order to evaluate the left-hand term of \eq{Mermin}. Note that each measured ``spin'' $s_{a,b}$ may independently be an integer or half an odd integer. The probability distribution is
\begin{align}
& P(s_a,m_a,s_b,m_b;\alpha,\beta) \nonumber\\
& = \tr{}{\rho' \(e^{-i\alpha S_y}\ket{s_a\ m_a}\bra{s_a\ m_a}e^{i\alpha S_y}\)\otimes\( e^{-i\beta S_y}\ket{s_b\ m_b}\bra{s_b\ m_b}e^{i\beta S_y}\)} \\
 & = 
 \sum_{s,s'}\tau_{sr}^2\tau_{s'r}^2\sum_{m,m'} (-1)^{s-m+s'-m'}\sum_{\sigma_a\ \mu_a}\sum_{\sigma_b\ \mu_b} 
\nonumber\\
& 
\eta_{a1}^{\sigma_a+\mu_a+\frac{s'-s+m'-m}2}\eta_{a2}^{\sigma_a-\mu_a+\frac{s'-s-m'+m}2} 
\eta_{b1}^{\sigma_b+\mu_b+\frac{s'-s+m'-m}2}\eta_{b2}^{\sigma_b-\mu_b+\frac{s'-s-m'+m}2}
\nonumber\\
& 
(1-\eta_{a1})^{-\sigma_a-\mu_a+s+m} (1-\eta_{a2})^{-\sigma_a+\mu_a+s-m}
(1-\eta_{b1})^{-\sigma_b-\mu_b+s+m} (1-\eta_{b2})^{-\sigma_b+\mu_b+s-m}  
\nonumber\\
&
\binom {s+m}{\sigma_a+\mu_a}^\frac12
\binom {s'+m'}{\sigma_a+\mu_a+s'-s+m'-m}^\frac12
\binom {s-m}{\sigma_a-\mu_a}^\frac12
\binom {s'-m'}{\sigma_a-\mu_a+s'-s-m'+m}^\frac12 
\nonumber\\
&
\binom {s+m}{\sigma_b+\mu_b}^\frac12
\binom {s'+m'}{\sigma_b+\mu_b+s'-s+m'-m}^\frac12
\binom {s-m}{\sigma_b-\mu_b}^\frac12
\binom {s'-m'}{\sigma_b-\mu_b+s'-s-m'+m}^\frac12 
\nonumber\\
&\bra {\sigma_a+{s'-s}, \mu_a+{m'-m}}e^{-i\alpha S_y}\ket{s_a, m_a}
\bra{s_a, m_a}e^{i\alpha S_y}\ket {\sigma_a, \mu_a} 
\nonumber\\
&\bra {\sigma_b+{s'-s}, \mu_b+{m'-m}}e^{-i\beta S_y}\ket{s_b, m_b}
\bra{s_b, m_b}e^{i \beta S_y}\ket {\sigma_b, \mu_b} 
\end{align}
We now use the angular momentum selection rules for rotation matrix elements,
%$\bra{s\ m} e^{-i\alpha S_y}\ket{s'\ m'}$ $\propto$ $\delta_{s,s'}$, 
which yields 
\begin{align}
\sigma_a+{s'-s} & = \sigma_a= s_a  \\
 \sigma_b+{s'-s} &= \sigma_b  = s_b 
\end{align}
If $s=s'$, this gives $\sigma_{a,b}=s_{a,b}$, respectively. However, if $s\neq s'$, no suitable $\sigma_{a,b}$ exists, which considerably simplifies the expression by introducing a $\delta_{s,s'}$ term.  Note that $m'-m$ is always an integer. We obtain 
\begin{align}
&P(s_a,m_a,s_b,m_b;\alpha,\beta) = \nonumber\\
& \sum_s\tau_{sr}^4\sum_{m,m'} (-1)^{2s-m-m'}\sum_{\mu_a,\mu_b} 
\eta_{a1}^{s_a+\mu_a+\frac{m'-m}2}\eta_{a2}^{s_a-\mu_a+\frac{-m'+m}2} 
\eta_{b1}^{s_b+\mu_b+\frac{m'-m}2}\eta_{b2}^{s_b-\mu_b+\frac{-m'+m}2}
\nonumber\\
& 
(1-\eta_{a1})^{-s_a-\mu_a+s+m} (1-\eta_{a2})^{-s_a+\mu_a+s-m}
(1-\eta_{b1})^{-s_b-\mu_b+s+m} (1-\eta_{b2})^{-s_b+\mu_b+s-m}  
\nonumber\\
&
\binom {s+m}{s_a+\mu_a}^\frac12
\binom {s+m'}{s_a+\mu_a+m'-m}^\frac12
\binom {s-m}{s_a-\mu_a}^\frac12
\binom {s-m'}{s_a-\mu_a-m'+m}^\frac12 
\nonumber\\
&
\binom {s+m}{s_b+\mu_b}^\frac12
\binom {s+m'}{s_b+\mu_b+m'-m}^\frac12
\binom {s-m}{s_b-\mu_b}^\frac12
\binom {s-m'}{s_b-\mu_b-m'+m}^\frac12 
\nonumber\\
&
\bra {s_a, \mu_a+{m'-m}}e^{-i\alpha S_y}\ket{s_a, m_a}
\bra{s_a, m_a}e^{i\alpha S_y}\ket {s_a, \mu_a} 
\nonumber\\
&\bra {s_b, \mu_b+{m'-m}}e^{-i\beta S_y}\ket{s_b, m_b}
\bra{s_b, m_b}e^{i \beta S_y}\ket {s_b, \mu_b}. 
\end{align}
This expression may already be evaluated, but for simplicity we will focus on the particular case where all losses are equal $\eta_j=\eta$, $\forall j$. 
 In that case we get the slightly simpler expression
\begin{align}
& P(s_a,m_a,s_b,m_b;\alpha,\beta) = 
\nonumber\\
& \sum_s\tau_{sr}^4\sum_{m,m'} (-1)^{2s-m-m'}\sum_{\mu_a,\mu_b} 
\eta^{2(s_a+s_b)} (1-\eta)^{2(2s-s_a-s_b)} 
\nonumber\\
&
\binom {s+m}{s_a+\mu_a}^\frac12
\binom {s+m'}{s_a+\mu_a+m'-m}^\frac12
\binom {s-m}{s_a-\mu_a}^\frac12
\binom {s-m'}{s_a-\mu_a-m'+m}^\frac12 
\nonumber\\
&
\binom {s+m}{s_b+\mu_b}^\frac12
\binom {s+m'}{s_b+\mu_b+m'-m}^\frac12
\binom {s-m}{s_b-\mu_b}^\frac12
\binom {s-m'}{s_b-\mu_b-m'+m}^\frac12 
\nonumber\\
&
d^{s_a}_{\mu_a+{m'-m},m_a}(\alpha)\ d^{s_a}_{\mu_a,m_a}(\alpha)\ 
d^{s_b}_{\mu_b+{m'-m},m_b}(\beta)\ d^{s_b}_{\mu_b,m_b}(\beta),
\label{eq:lhs}
\end{align}
with the well-known expression for the rotation matrix elements 
\begin{align}
d^s_{m_1, m_2}(\alpha) 
&=\bra{s\ m_1}e^{-i\alpha S_y}\ket{s\ m_2} 
\nonumber\\
& =  \sqrt{\frac{(s + m_1)!(s - m_1)!}{(s + m_2)!(s - m_2)!}} \(\cos \frac\alpha2\)^{m_1 + m_2}
\(\sin \frac\alpha2\)^{m_1 - m_2}  P^{(m_1 - m_2, m_1 + m_2)}_{s - m_1}(\cos\alpha),
\end{align}
where the last symbol denotes a Jacobi polynomial. Note that $\eta=1$ forces $2s-s_a-s_b=0$ and, in fact, $s=s_a=s_b$ since the photon numbers and therefore, spin magnitudes, are perfectly correlated between Alice and Bob. This is of course not the case when $\eta<1$ as photons are randomly lost between the various detectors. The ideal efficiency should of course give the same result as  Mermin's original inequality, which we will verify numerically in the next section. 

%Let's check this.
%\begin{widetext}
%\begin{align}
%P = & \(\frac{\tanh^s r}{\cosh r}\)^4\sum_{m,m'} (-1)^{2s-m-m'}\sum_{\mu_a,\mu_b} \nonumber\\
%&
%\binom {s+m}{s+\mu_a+\frac{m-m'}2}^\frac12
%\binom {s+m'}{s+\mu_a+\frac{m'-m}2}^\frac12
%\binom {s-m}{s-\mu_a+\frac{-m+m'}2}^\frac12
%\binom {s-m'}{s-\mu_a+\frac{-m'+m}2}^\frac12 \nonumber\\
%&
%\binom {s+m}{s+\mu_b+\frac{m-m'}2}^\frac12
%\binom {s+m'}{s+\mu_b+\frac{m'-m}2}^\frac12
%\binom {s-m}{s-\mu_b+\frac{-m+m'}2}^\frac12
%\binom {s-m'}{s-\mu_b+\frac{-m'+m}2}^\frac12\nonumber\\
%&\bra {s\ \mu_a+\frac{m'-m}2}e^{-i\alpha S_y}
%\Big|s\ m_a\Big\rangle\Big \langle s\ m_a\Big|
%e^{i\alpha S_y}\ket {s\ \mu_a+\frac{m-m'}2} \nonumber\\
%&\bra {s\ \mu_b+\frac{m'-m}2}e^{-i\beta S_y}
%\Big|s\ m_b\Big\rangle\Big \langle s\ m_b\Big|
%e^{i \beta S_y}\ket {s\ \mu_b +\frac{m-m'}2} \
%\end{align}
%\end{widetext}

\subsubsection{Crosscorrelations}

We now derive the expressions for the crosscorrelations on the right-hand side of \eq{Mermin}. The arbitrary spin component of \eq{alpha} can be written in terms of spin ladder operators
\begin{equation}
S_\alpha  = \cos\alpha\ S_z + \sin\alpha\ \frac{S_++S_-}2 
\end{equation}
which we can use in the crosscorrelation expression
\begin{equation}
\mathcal C_{\alpha\beta} = \left\langle S_{A,\alpha}\, S_{B,\beta}\right\rangle. 
\end{equation}
We then evaluate these matrix elements using the decohered entangled density operator $\rho'$.
\begin{align}
\mathcal C_{\alpha\beta} =\ & \tr{}{\rho' \(e^{i\alpha S_{Ay}}S_{Az}e^{-i\alpha S_{Ay}}\)\otimes \(e^{i\beta S_{By}}S_{Bz}e^{-i\beta S_{By}}\)} \\
=\ & \mathrm{Tr}\left[\rho' \(\cos\alpha\ S_{Az} +  \frac{\sin\alpha}2\ (S_{A+}+S_{A-})\)\right.\nonumber\\
 & \ \quad\otimes \left.\(\cos \beta\ S_{Bz} +  \frac{\sin \beta}2\ (S_{B+}+S_{B-})\)\right].
\end{align}
Again, we take all losses equal to simplify the computations. This yields
\begin{align}
\mathcal C_{\alpha\beta}= &\sum_{s=0}^\infty\tau_{sr}^4\sum_{m=-s}^s\left\{  %m'=m
(-1)^{2(s-m)}\cos\alpha\ \cos \beta 
\sum_{\sigma_a=\sigma_\mathit{min}}^s\sum_{\mu_a=-\sigma_a}^{\sigma_a}
\sum_{\sigma_b=\sigma_\mathit{min}}^s\sum_{\mu_b=-\sigma_b}^{\sigma_b} \right.
\nonumber\\
&
\qquad\qquad\qquad\qquad
\eta^{2(\sigma_a+ \sigma_b)} (1-\eta)^{2(-\sigma_a-\sigma_b +2s+2m)} 
\nonumber\\
&
\qquad\qquad\qquad\qquad
\binom {s+m}{\sigma_a+\mu_a}
\binom {s-m}{\sigma_a-\mu_a} 
\binom {s+m}{\sigma_b+\mu_b}
\binom {s-m}{\sigma_b-\mu_b}
\mu_a\ \mu_b
\nonumber\\ 
&    %m'=m+1
\qquad\qquad\qquad+(-1)^{2(s-m)-1}\frac{\sin\alpha\sin \beta}4 
\sum_{\sigma_a=\sigma_\mathit{min}}^s\sum_{\mu_a=-\sigma_a}^{\sigma_a}
\sum_{\sigma_b=\sigma_\mathit{min}}^s\sum_{\mu_b=-\sigma_b}^{\sigma_b}
\nonumber\\
&
\qquad\qquad\qquad\qquad
\eta^{2(\sigma_a+ \sigma_b +1)}
(1-\eta)^{2(-\sigma_a-\sigma_b +2s+2m)} 
\nonumber\\
&
\qquad\qquad\qquad\qquad
\binom {s+m}{\sigma_a+\mu_a}^\frac12
\binom {s+m+1}{\sigma_a+\mu_a+1}^\frac12
\binom {s-m}{\sigma_a-\mu_a}^\frac12
\binom {s-m-1}{\sigma_a-\mu_a-1}^\frac12 
\nonumber\\
&
\qquad\qquad\qquad\qquad
\binom {s+m}{\sigma_b+\mu_b}^\frac12
\binom {s+m+1}{\sigma_b+\mu_b+1}^\frac12
\binom {s-m}{\sigma_b-\mu_b}^\frac12
\binom {s-m-1}{\sigma_b-\mu_b-1}^\frac12 
\nonumber\\
&
\qquad\qquad\qquad\qquad
\sqrt{\sigma_a(\sigma_a+1)-\mu_a(\mu_a+1)}\sqrt{\sigma_b(\sigma_b+1)-\mu_b(\mu_b+1)}
%\bra {\sigma_a,\mu_a+1}  \ S_{A+} \ket {\sigma_a,\mu_a}_A
%\bra {\sigma_b,\mu_b+1}   \ S_{B+}\ket {\sigma_b,\mu_b}_B
\nonumber\\ 
&    %m'=m-1
\qquad\qquad\qquad+(-1)^{2(s-m)+1}\frac{\sin\alpha\sin \beta}4 
\sum_{\sigma_a=\sigma_\mathit{min}}^s\sum_{\mu_a=-\sigma_a}^{\sigma_a}
\sum_{\sigma_b=\sigma_\mathit{min}}^s\sum_{\mu_b=-\sigma_b}^{\sigma_b}
\nonumber\\
&
\qquad\qquad\qquad\qquad
\eta^{2(\sigma_a+ \sigma_b -1)}
(1-\eta)^{2(-\sigma_a-\sigma_b +2s+2m)} 
\nonumber\\
&
\qquad\qquad\qquad\qquad
\binom {s+m}{\sigma_a+\mu_a}^\frac12
\binom {s+m-1}{\sigma_a+\mu_a-1}^\frac12
\binom {s-m}{\sigma_a-\mu_a}^\frac12
\binom {s-m+1}{\sigma_a-\mu_a+1}^\frac12 
\nonumber\\
&
\qquad\qquad\qquad\qquad
\binom {s+m}{\sigma_b+\mu_b}^\frac12
\binom {s+m-1}{\sigma_b+\mu_b-1}^\frac12
\binom {s-m}{\sigma_b-\mu_b}^\frac12
\binom {s-m+1}{\sigma_b-\mu_b+1}^\frac12 
\nonumber\\
&
\qquad\qquad\qquad\qquad
\sqrt{\sigma_a(\sigma_a+1)-\mu_a(\mu_a-1)}\sqrt{\sigma_b(\sigma_b+1)-\mu_b(\mu_b-1)}\biggr\}
%\bra {\sigma_a,\mu_a-1}\frac{\sin\alpha}2\ S_{A-} \ket {\sigma_a,\mu_a}_A
%\bra {\sigma_b,\mu_b-1}   \frac{\sin \beta}2\ S_{B-}\ket {\sigma_b,\mu_b}_B
\label{eq:rhs}
\end{align}
We are now ready to  make use of \eqs{lhs}{rhs} to evaluate Mermin's inequality quantum mechanically as a function of the loss parameter, or detection efficiency, $\eta$. 

\section{Calculation results}

We evaluated \eqs{lhs}{rhs} numerically for a range of detector angles $(\alpha,\beta)$, efficiencies $\eta$, and squeezing parameters $r$. After defining an angle $\alpha-\pi/2=\beta-\pi/2=\theta$ as Mermin does, our expressions reproduce Mermin's results exactly at perfect efficiency $\eta=1$. 
%small squeezing simply means that large photon numbers are produced much less often than small ones. 
Evaluation of these probabilities at perfect efficiency is straightforward because the sum over variable $s$ reduces to a single term, as the $(1-\eta)^{2(2s-s_a-s_b)}$ term goes to zero for all but the single spin in question. In this regime, the experimental results are exactly equivalent to those achieved by Stern-Gerlach style measurements on a spin $s$ particle.

In the case $\eta<1$, the sum in the probability over variable $s$ cannot be neglected. To evaluate these expressions, we cut off the sum over $s$ at a point determined dynamically in the computation, depending on the squeezing and the efficiency. Then, we carried the sum out further to verify that our results converged to good precision. The numerical evaluation of these expressions placed heavy demands on computational time, especially for high-squeezing, low-efficiency, or high-spin systems, all of which required summing over many values for $s$. For these difficult computations, we made use of the ``Dogwood'' parallel computing cluster managed by the University of Virginia Alliance for Computational Science and Engineering. Most of these expressions were evaluated on 1-4 computing nodes, each equipped with two four-core, 3 GHz Intel Xeon EMT processors with 3 GB of RAM. Typical run time was anywhere from several minutes for the least demanding jobs to over 150 hours for the most demanding.

\subsection{Influence of decoherence on spin analyzer angles}

\begin{figure} [htbp]
%\vspace*{13pt}
%\centerline{\epsfig{file=Spin1Theta, width=8.2cm}} %100 percent
\centerline{\includegraphics[width=4.in]{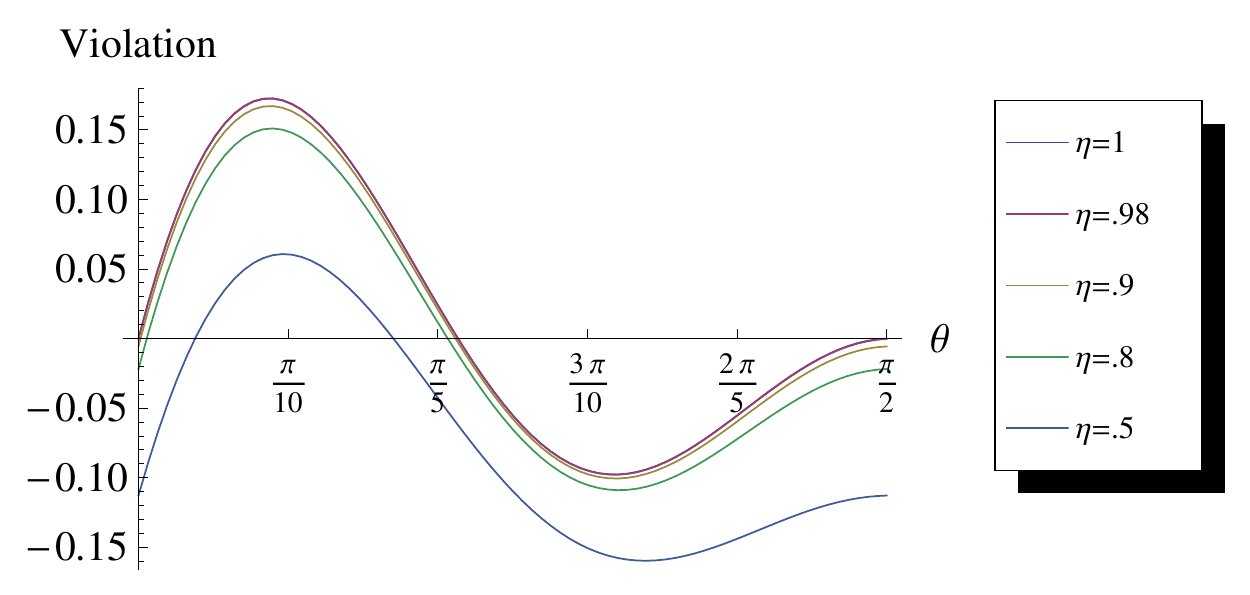}}
\vspace*{13pt}
\caption{\label{fig:SmallSpinTheta}Plot of violation versus angle for different efficiencies for the $s=1$ case. Higher values constitute greater violation; negative results can be explained by a local hidden variable model. Decreasing efficiencies produce decreasing violation. Here, $r=0.5$.}
\end{figure}

\begin{figure} [htbp]
%\vspace*{13pt}
%\centerline{\epsfig{file=Spin2Theta, width=8.2cm}} %100 percent
\centerline{\includegraphics[width=4.in]{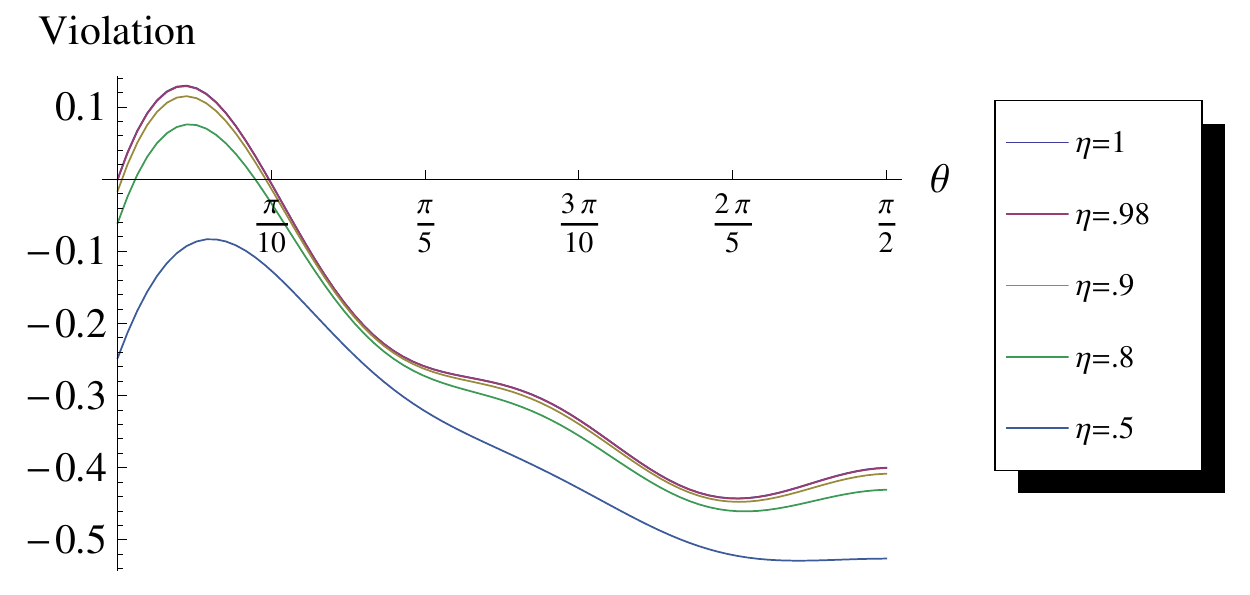}}
\vspace*{13pt}
\caption{\label{fig:BigSpinTheta}Plot of violation versus angle for different efficiencies for the $s=2$ (four photon) case. The results are much the same as in \fig{SmallREtaPlot}, except the magnitude of violation is lower. Note that there is also a smaller angular range of violation and a greater sensitivity to inefficiencies. Again, $r=0.5$.}
\end{figure}

Figures \ref{fig:SmallSpinTheta} and \ref{fig:BigSpinTheta} show the violation of Mermin's inequality, which is defined as the difference of the right-hand and left-hand sides of \eq{Mermin}, for different efficiencies $\eta$ at fixed spin $s$ and OPO coefficient $r$. The inequality is violated when the ``Violation'' is positive in \figs{SmallSpinTheta}{BigSpinTheta}. The efficiencies decrease from the top to the bottom of the plots---higher efficiencies violate over a greater range of angles, and higher spins require higher efficiencies to produce violations.

%%%%%%%%%%%%%%%%%%%%%%%%%%%%%%%%%%%%%%%%%%%%%%%%%%%%%%%%%%%%%%%%%%
%%%%%%%%%%%%%%%%%%%%%%%%%%%%%%%%%%%%%%%%%%%%%%%%%%%%%%%%%%%%%%%%%%

%%%%%%%%%%%%%%%%%%%%%%%%%%%%%%%%%%%%%%%%%%%%%%%%%%%%%%%%%%%%%%%%%%
%%%%%%%%%%%%%%%%%%%%%%%%%%%%%%%%%%%%%%%%%%%%%%%%%%%%%%%%%%%%%%%%%%

Unsurprisingly, a lower efficiency $\eta$ results in considerably reduced violation margins for Mermin's inequality; as detailed below, violation also decreases with an increase in either the squeezing parameter $r$ or the spin $s$.  Additionally, the optimal angle for violation appears to shift very slightly as the efficiency decreases.

It's also important to recall that the situation considered here is perfectly symmetric. All of the detectors have equal efficiency, and the choice of optimal angle $\theta$ (defined by Mermin \cite{Mermin1980}) depends crucially on the equivalence of the two detectors: the otherwise independent detector angles $\alpha$ and $\beta$ are identical in this model. Further optimizing the violation of Mermin's inequality over the (really four-dimensional) parameter space of the angles might yield better results for a particular choice of experimental conditions. In the results that follow, however, we only evaluated the inequality at the optimal angle for perfect detection. Thus, the violations we predicted here are, at worst, lower bounds for actual violations.

\subsection{Influence of decoherence on violation for different spins $s$}

\begin{figure} [htbp]
%\vspace*{13pt}
%\centerline{\epsfig{file=fig1, width=8.2cm}} %100 percent
\centerline{\includegraphics[width=4.in]{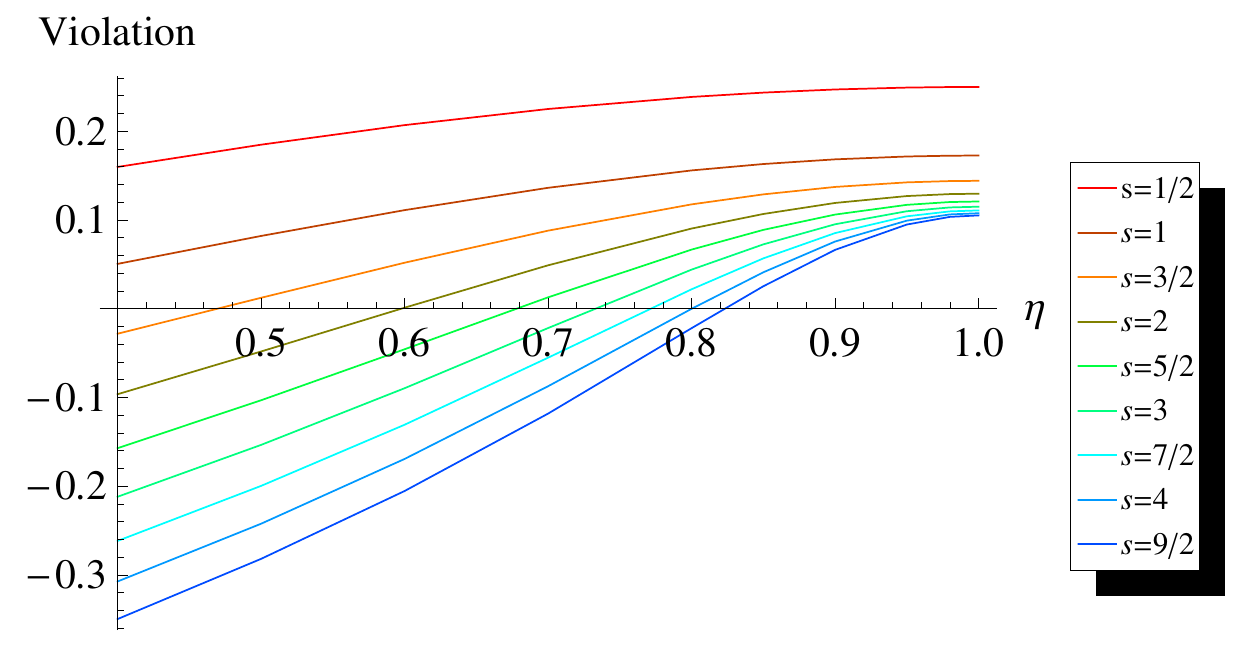}}
\vspace*{13pt}
\caption{\label{fig:BigREtaPlot}Plot of violation versus efficiency for different spins $s$. Again, higher values constitute greater violation. Decreasing efficiencies produce decreasing violation, and lower spins violate more strongly than higher spins. Here, $r=0.4$.}
\end{figure}

\begin{figure} [htbp]
%\vspace*{13pt}
%\centerline{\epsfig{file=fig1, width=8.2cm}} %100 percent
\centerline{\includegraphics[width=4.in]{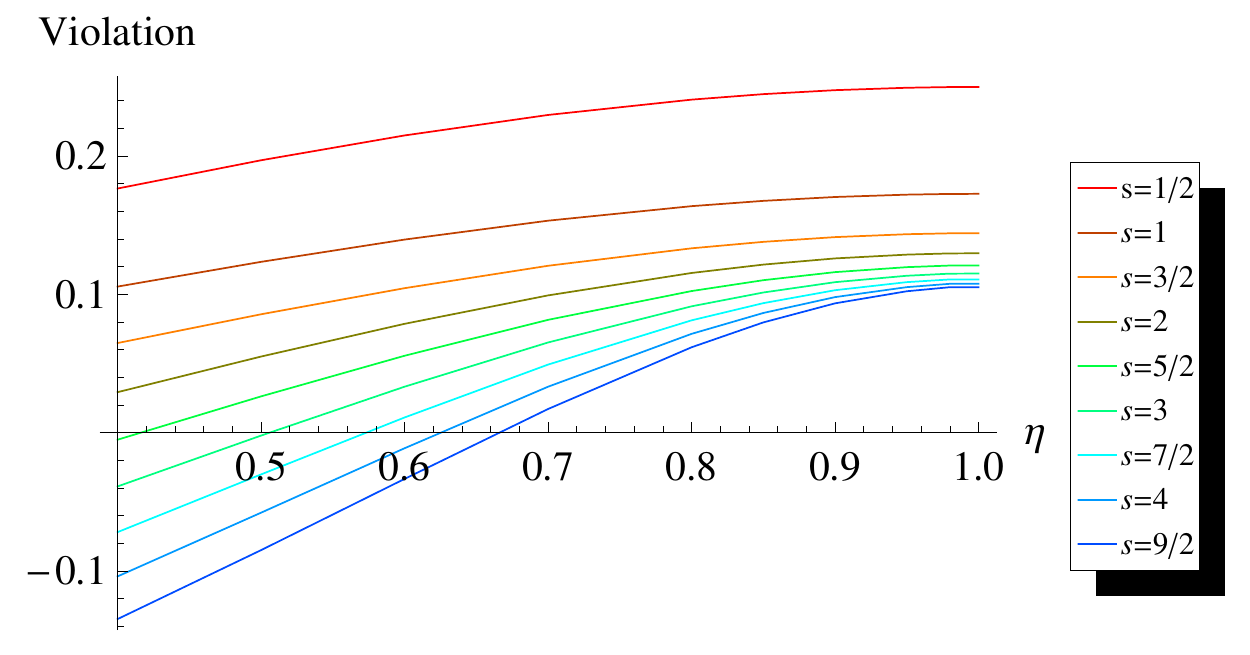}}
\vspace*{13pt}
\caption{\label{fig:SmallREtaPlot}The same plot as \fig{BigREtaPlot} for $r=0.2$. Note that the general trends are the same, but violation is stronger. Generally, higher values for $r$ give lower violation for $\eta<1$ and vice versa.}
\end{figure}

After choosing to evaluate these results at the optimal angles for $\eta=1$, we can examine the effect of inefficiencies more directly by plotting the violation against the efficiency, for $r=0.4$, \fig{BigREtaPlot}, and $r=0.2$, \fig{SmallREtaPlot}. These plots are  constructed by taking a single point (the maximum in the case of perfect efficiency, near the maximum for lower efficiencies) and combining these points in a plot of violation vs. $\eta$. Different lines on the same plot represent, from top to bottom, different spins from $s={1}/{2}$ to $s={9}/{2}$ in steps of ${1}/{2}$. 

%%%%%%%%%%%%%%%%%%%%%%%%%%%%%%%%%%%%%%%%%%%%%%%%%%%%%%%%%%%%%%%%%%
%%%%%%%%%%%%%%%%%%%%%%%%%%%%%%%%%%%%%%%%%%%%%%%%%%%%%%%%%%%%%%%%%%

%%%%%%%%%%%%%%%%%%%%%%%%%%%%%%%%%%%%%%%%%%%%%%%%%%%%%%%%%%%%%%%%%%
%%%%%%%%%%%%%%%%%%%%%%%%%%%%%%%%%%%%%%%%%%%%%%%%%%%%%%%%%%%%%%%%%%

As \figs{BigREtaPlot}{SmallREtaPlot} show, as the efficiency decreases, all but the lowest spins fail to violate the inequality; the higher the spins, the more deleterious the effect of inefficiency on violation. In \fig{SmallREtaPlot}, though, all spins tend to violate more strongly than in  \fig{BigREtaPlot}, especially at low efficiencies. This trend is general---the lower the squeezing parameter $r$, the less effect inefficiency seems to have on violating the inequality.

\subsection{Joint influences of decoherence and squeezing}

\begin{figure} [htbp]
%\vspace*{13pt}
%\centerline{\epsfig{file=fig1, width=8.2cm}} %100 percent
\centerline{\includegraphics[width=4.in]{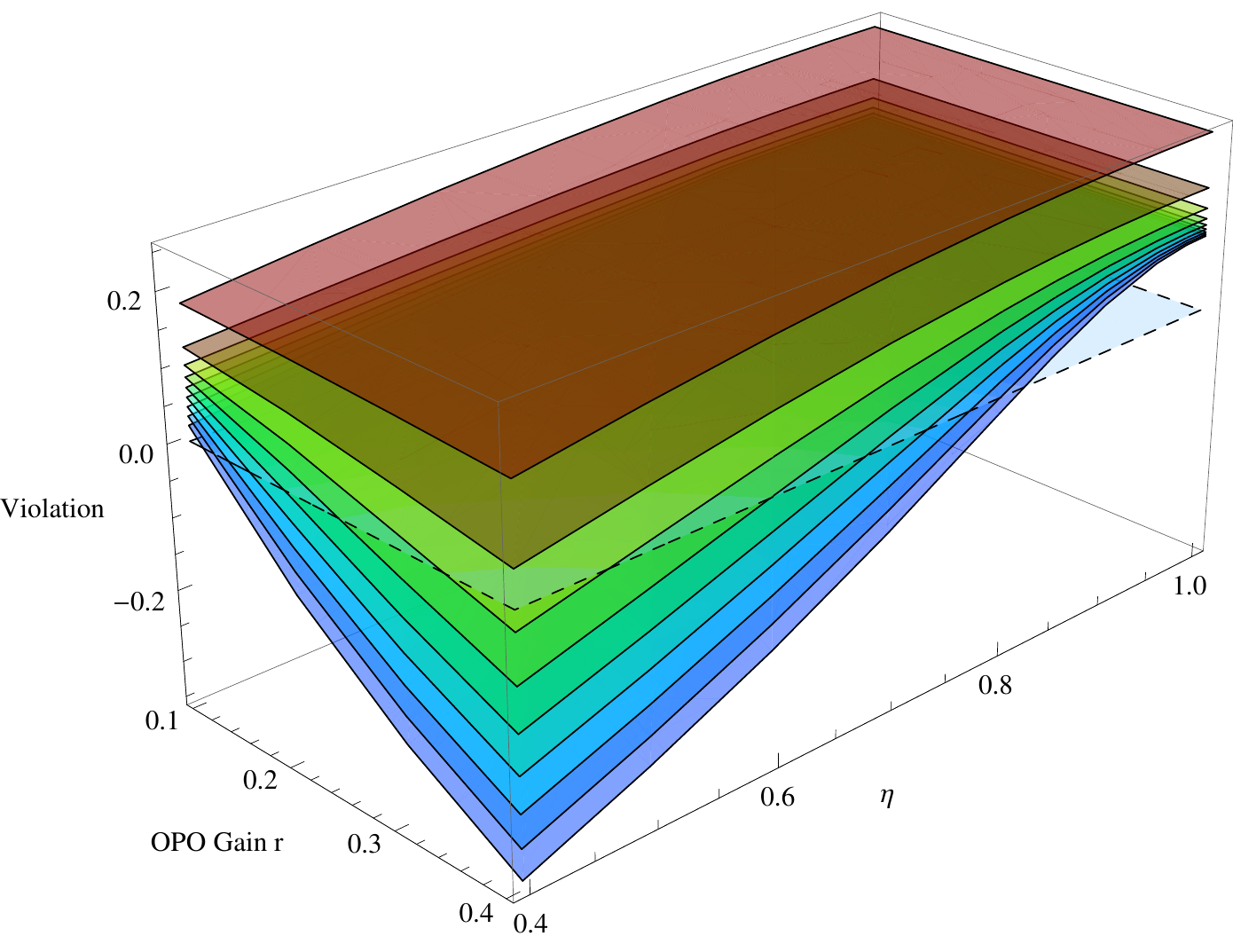}}
\vspace*{13pt}
\caption{\label{fig:rainbow}This plot compares the effect of both squeezing $r$ and efficiency $\eta$ on the extent of violation. The nine surfaces show the predicted violation for the nine different spins. The top surface is that for $s=\frac{1}{2}$. The next surface is that for $s=1$, and so on, through $s={9}/{2}$. Note that the surfaces are labeled with the same color scheme as in \fig{SmallREtaPlot} and \fig{BigREtaPlot}---these two plots are just the cross sections through the two planes $r=0.2$ and $r=0.4$ in this figure. As a reference, a plane is also plotted at zero violation.}
\end{figure}

Many of our results can be summarized in the plot of \fig{rainbow}, which again shows the dependence of the violation of Mermin's inequality on $r$ and $\eta$. Note that \figs{BigREtaPlot}{SmallREtaPlot} are cross-sections of the \fig{rainbow} at constant $r$, and that \figs{SmallSpinTheta}{BigSpinTheta} have their maxima represented as single points on \fig{rainbow}. Crucially, violations of Mermin's inequality are predicted for all of these spins under a range of experimentally attainable conditions. 

%%%%%%%%%%%%%%%%%%%%%%%%%%%%%%%%%%%%%%%%%%%%%%%%%%%%%%%%%%%%%%%%%%
%%%%%%%%%%%%%%%%%%%%%%%%%%%%%%%%%%%%%%%%%%%%%%%%%%%%%%%%%%%%%%%%%%

There are several trends in these data. First, the extent of violation of the inequality decreases monotonically as the detection efficiency decreases. This was expected. The extent of the decrease, however, depends both on the amount of squeezing and on the total spin itself. On the one hand, low spins almost always violate under the range of conditions plotted here, and have little dependence on either parameter ($\eta$ or $r$). Higher spin systems violate less strongly---even in the ideal case of perfect efficiency---and the extent of violation decreases much more rapidly with $\eta$ than for lower spin systems. That exacerbated sensitivity of high-spin quantum systems to decoherence is not too surprising either. However, it is important to mention here that several inequalities since Mermin's time make violation more robust by remedying one or more of these issues. Garg and Mermin later produced an inequality whose angular range of violation does not decrease with increasing spin \cite{Garg1982}, and more recent work has produced simple inequalities that are violated even for arbitrarily large spin \cite{Peres1992, Kaszlikowski2002, Collins2002} approaching the maximum possible violation for a $2\times 2\times d$ Bell-type experiment  \cite{Landau1987, Durt2001,Cirelson1980}. The CGLMP inequality \cite{Collins2002} in particular, produces large violations even for large spins and also appears to be resistant to noise in the form of Werner states. We've begun exploring the CGLMP inequality in our optical framework, but the form of the inequality makes incorporating loss more computationally intensive than in Mermin's 1980 inequality. In future work, we plan to address the optical realization of these newer high-spin inequalities.

\begin{figure} [htbp]
%\vspace*{13pt}
\centerline{\includegraphics[width=4.in]{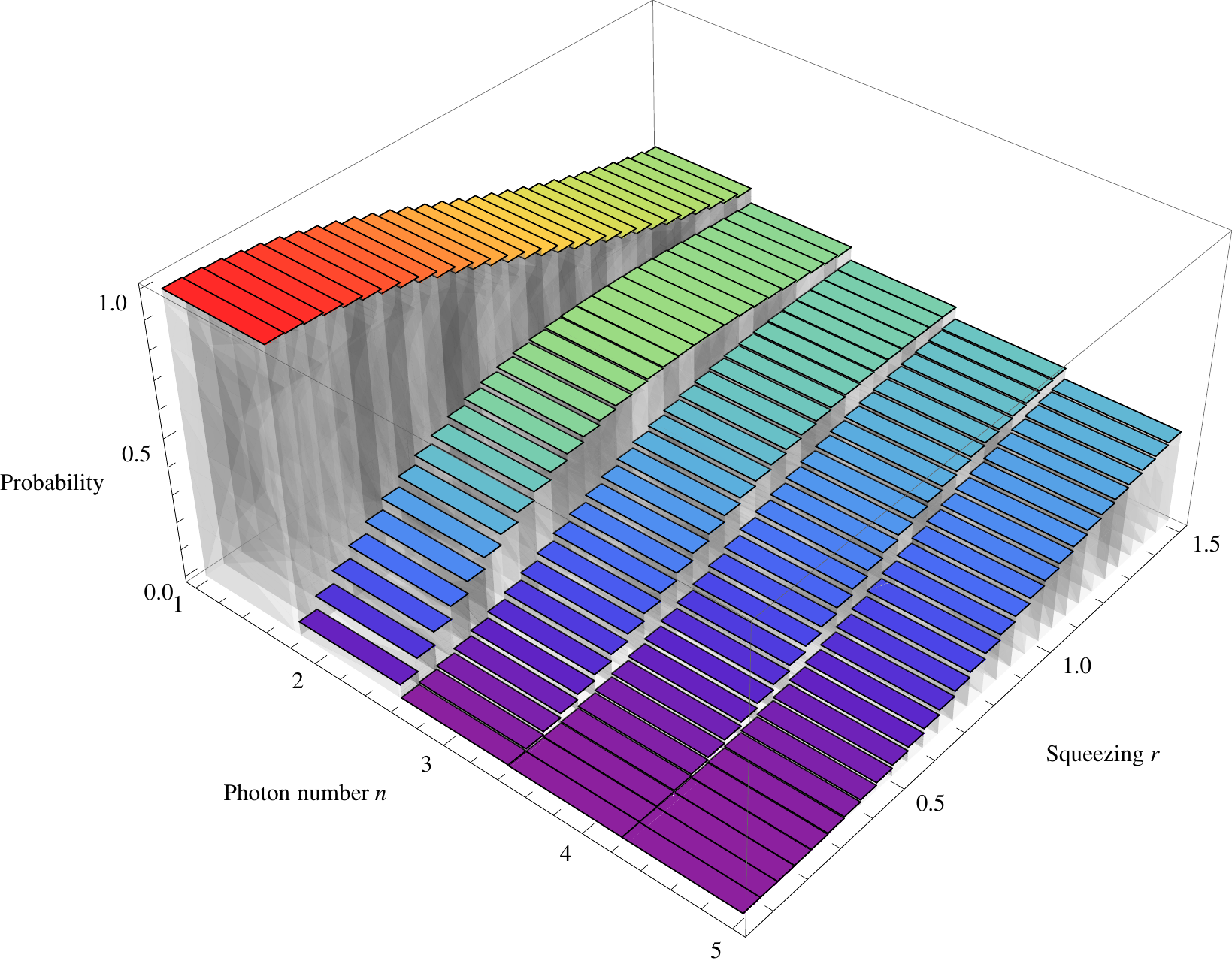}}
\vspace*{13pt}
\caption{\label{fig:sqdist}Plot of the relative probability of producing a certain photon number in a Fock state as a function of the squeezing $r$. As the squeezing increases, generating larger and larger photon number states becomes more and more probable.}
\end{figure}

Finally, we see from \fig{rainbow} that a given spin $s$ produced by a strongly squeezed source is much more sensitive to decoherence than the same spin $s$ produced by a weakly squeezed source. In the case of perfect detectors, the amount of squeezing doesn't affect the extent of the violations, but a weakly squeezed state actually produces greater violations than a highly squeezed state if the efficiency is low. This is a much less intuitive result if we associate a more highly squeezed state with less classical behavior, and thus expect a more highly squeezed state to produce greater violations. Our result can be simply explained by the fact that the photon-number distribution of the output fields depends strongly on the squeezing parameter: in EPR states as produced by the OPO, the probability amplitude of a given Fock state is $\tau_{s,r}=(\tanh r)^n/\cosh r$ , which in both limits goes as a decaying exponential in $r$. So, if the OPO is barely squeezing, the probability for generating a total photon number, or spin state $s$, is much larger than the probability of generating even the next larger spin, whereas, as $r$ increases, all of the photon-number states become equally probable, as displayed in \fig{sqdist}.
%%%%%%%%%%%%%%%%%%%%%%%%%%%%%%%%%%%%%%%%%%%%%%%%%%%%%%%%%%%%%%%%%%
%%%%%%%%%%%%%%%%%%%%%%%%%%%%%%%%%%%%%%%%%%%%%%%%%%%%%%%%%%%%%%%%%%
Inefficiencies thus effect a sort of ``crossover'' between the photon-number states, with a strength increasing as the inefficiency increases: the beamsplitters that model losses annihilate photons from the signal field, which reduces the total spin corresponding to that Fock state in the Schwinger representation. Then, because the measured results are partitioned into individual spin-$s$ subspaces before checking for violation, the measured distribution of projections in a given subspace depends both on the distribution in that subspace and on the distribution in higher subspaces, because the higher subspaces will couple to the lower subspaces with an interaction strength determined by the beamsplitter reflection coefficient, or inefficiency of the detector. The relative amplitudes of the two subspaces are therefore crucially important: on the one hand, even if there is strong coupling between the two subspaces through an inefficient detector, if the lower spin subspace has a significantly higher associated probability amplitude, then the effect of the higher spin subspace will be negligible, and this is what happens in the low-squeezing case. If, on the other hand, there is high squeezing, then the two spin subspaces have comparable probability amplitudes and there is a large effect if the detectors are inefficient. For imperfect detectors, higher squeezing thus causes many of the same effects as lower efficiency because both cause a lower-spin subspace to have increased coupling to a higher spin subspace.

Nonetheless, it's important to remember that there is a significant practical difference between increased squeezing and decreased efficiency. Although theoretically, lower squeezing produces higher violations given inefficient detectors, with low squeezing the OPA produces high photon number states much less frequently. So, there's a tradeoff: low squeezing will produce higher violations than high squeezing in a nonideal detection case, but one will have to wait much longer to see the violations in an experiment. Increasing the efficiency allows for increased squeezing, which is crucially important for violating any high-spin inequality in a reasonable experimental timeframe.

\section{Conclusion}
Our results show that it appears quite feasible to violate Mermin's high-spin Bell inequality under a range of experimentally viable conditions. After converting the photon number distribution produced by an optical parametric oscillator from the Fock basis to the spin basis, we find that the state could be written as combination of singlet states typically used to violate Bell-type inequalities. Incorporating inefficiencies into our probabilities by considering the effect of loss through beamsplitters allows us to determine the effect of actual experimental parameters on the outcome of our tests.  In particular, we find that Mermin's inequality is violated for $s={1}/{2}$ to $s={9}/{2}$ under a wide range of experimentally realizable efficiencies and squeezing parameters. 

In the future, we hope to build on these results by performing an experimental violation of these inequalities using the setup described previously. We hope also to examine other high-dimensional inequalities such as that of Collins, Gisin, Linden, Massar, and Popescu\cite{Collins2002} which---though more computationally intensive in this framework---may offer greater violations and increased resistance to experimental imperfections like loss and noise. 

This work was supported by NSF grants PHY-0960047 and PHY-0855632, and by the FINESSE program of Raytheon BBN, under the DARPA ``Information in a Photon'' program. RE was partially supported by an Echols Scholar Ingrassia Research grant  at the University of Virginia.

%\bibliographystyle{apsrev4-1.bst}

%\bibliographystyle{/Users/opfister/Documents/Werk/Articles/Authors/P/Pfister/BibStyles/bibstyleNCM}
%\bibliography{/Users/opfister/Documents/Werk/Articles/Authors/P/Pfister/Pfister}

\end{document}